\documentclass{aa}
\usepackage{epsfig}
\usepackage{graphicx}
\usepackage{txfonts}

\begin{document}

\title{AGILE observation of a gamma-ray flare from the blazar 3C 279}

  \author{A.\ Giuliani   \inst{1,*} %
     \and F.\ D'Ammando   \inst{2,3}
     \and S.\ Vercellone    \inst{1}   
     \and V.\ Vittorini   \inst{2,4}  
     \and A.\ W.\ Chen    \inst{1,4}
     \and I.\ Donnarumma  \inst{2}
     \and L.\ Pacciani    \inst{2}
     \and G.\ Pucella     \inst{2}
     \and A.\ Trois       \inst{2}   
     \and A.\ Bulgarelli  \inst{5}
     \and F.\ Longo       \inst{6}    
     \and M.\ Tavani      \inst{2,3}
     \and G.\ Tosti        \inst{7}
     \and D.\ Impiombato   \inst{7}    
     \and A.\ Argan       \inst{2}
     \and G.\ Barbiellini \inst{6}
     \and F.\ Boffelli    \inst{8,9}
     \and P.\ A.\ Caraveo \inst{1}
     \and P.\ W.\ Cattaneo \inst{8}
     \and V.\ Cocco       \inst{2}
     \and E.\ Costa       \inst{2}
     \and E.\ Del Monte   \inst{2}
     \and G.\ De Paris    \inst{2}
     \and G.\ Di Cocco    \inst{5}
     \and Y.\ Evangelista \inst{2}
     \and M.\ Feroci      \inst{2}
     \and M.\ Fiorini     \inst{1}
     \and F.\ Fornari     \inst{1}
     \and T.\ Froysland   \inst{3,4}
     \and F.\ Fuschino    \inst{5}
     \and M.\ Galli       \inst{10}
     \and F.\ Gianotti    \inst{5}
     \and C.\ Labanti     \inst{5}
     \and Y.\ Lapshov     \inst{2}
     \and F.\ Lazzarotto  \inst{2}
     \and P.\ Lipari      \inst{11}
     \and M.\ Marisaldi   \inst{5}
     \and S.\ Mereghetti  \inst{1}
     \and A.\ Morselli    \inst{12}
     \and A.\ Pellizzoni  \inst{1}
     \and F.\ Perotti     \inst{1}
     \and P.\ Picozza     \inst{12}
     \and M.\ Prest       \inst{13}
     \and M.\ Rapisarda   \inst{14}
     \and A.\ Rappoldi    \inst{8}
     \and P.\ Soffitta    \inst{2}
     \and M.\ Trifoglio   \inst{5}
     \and E.\ Vallazza    \inst{6}
     \and A.\ Zambra      \inst{1}
     \and D.\ Zanello     \inst{9}  
     \and S.\ Cutini      \inst{15}
     \and D.\ Gasparrini  \inst{15}
     \and C.\ Pittori     \inst{15}
     \and B.\ Preger      \inst{15}
     \and P.\ Santolamazza \inst{15} 
     \and F.\ Verrecchia  \inst{15} 
     \and P.\ Giommi      \inst{15}
     \and S.\ Colafrancesco \inst{15}
     \and L.\ Salotti     \inst{16}
    }
   \institute{INAF/IASF--Milano, Via E.\ Bassini 15, I-20133 Milano, Italy            
   \and INAF/IASF--Roma, Via Fosso del Cavaliere 100, I-00133 Roma, Italy
   \and Dip. di Fisica, Univ. ``Tor Vergata'', Via della Ricerca
  Scientifica 1, I-00133 Roma, Italy
   \and CIFS--Torino, Viale Settimio Severo 3, I-10133, Torino, Italy
   \and INAF/IASF--Bologna, Via Gobetti 101, I-40129 Bologna, Italy
   \and Dip. di Fisica and INFN, Via Valerio 2, I-34127 Trieste, Italy
  \and Dip. di Fisica, Univ. di Perugia, Via Pascoli, I-06123 Perugia, Italy
   \and INFN--Pavia, Via Bassi 6, I-27100 Pavia, Italy
   \and Dip. Fisica Nucleare e Teorica, Univ. di Pavia, Via Bassi 6, I-27100
   Pavia, Italy
   \and ENEA--Bologna, Via dei Martiri di Monte Sole 4, I-40129 Bologna, Italy  
   \and INFN--Roma ``La Sapienza'', Piazzale A. Moro 2, I-00185 Roma, Italy
   \and INFN--Roma ``Tor Vergata'', Via della Ricerca Scientifica 1, I-00133 Roma, Italy
   \and Dip. di Fisica, Univ. dell'Insubria, Via Valleggio 11, I-22100 Como, Italy
   \and ENEA--Roma, Via E. Fermi 45, I-00044 Frascati (Roma), Italy
   \and ASI--ASDC, Via G. Galilei, I-00044 Frascati (Roma), Italy
   \and ASI, Viale Liegi 26 , I-00198 Roma, Italy
}

\offprints{giuliani@iasf-milano.inaf.it, *Corresponding Author: Andrea Giuliani} 
\date{received; accepted}
\abstract{We report the detection by the AGILE satellite of an intense
 gamma-ray flare from the gamma-ray source 3EG J1255-0549, associated to
 the Flat Spectrum Radio Quasar 3C 279, during the AGILE pointings
 towards the Virgo Region on 2007 July 9-13.}
{The simultaneous optical, X-ray and gamma-ray covering
allows us to study the spectral energy distribution (SED) and the theoretical models
relative to the flaring episode of mid-July.}
{AGILE observed the source during its Science Performance Verification Phase with its two co-aligned imagers: the Gamma-Ray
Imaging Detector (GRID) and the hard X-ray imager (Super-AGILE) sensitive in the 30 MeV -- 50 GeV and 18 -- 60 keV respectively. 
During the AGILE observation the source was
 monitored simultaneously in optical band by the REM telescope and in the
 X-ray band by the \textit{Swift} satellite through 4 ToO observations.}  
{During 2007 July 9-13 July 2007, AGILE-GRID detected gamma-ray emission from 3C 279,
with the source at $\sim$ 2$^{\circ}$ from the center of the Field of View,
 with an average flux of (210 $\pm$ 38)  $\times$ 10$^{-8}$ ph cm$^{-2}$
 s$^{-1}$ for energy above 100 MeV. No emission was detected by Super-AGILE,
 with a 3-$\sigma$ upper limit of 10 mCrab. During the observation lasted
 about 4 days no significative gamma-ray flux variation was observed.}
{The Spectral Energy Distribution is modelled with a homogeneous one-zone
Synchrotron Self Compton emission plus the contributions by external Compton scattering of direct disk radiation
and, to a lesser extent, by external Compton scattering of photons  
from the Broad Line Region. }

\keywords{gamma rays: observations -- galaxies: quasars: general -- galaxies: quasars: individual 3C 279}
\authorrunning{A. Giuliani et al.}
\titlerunning{AGILE observation of a gamma-ray flare from 3C 279}
\maketitle

\section{Introduction}
3C 279 (z=0.536) is an optically violent variable (OVV) quasar, the
first and one of the brightest blazar discovered to emit in gamma-ray band by
EGRET (Hartman et al. 1992). 

Despite its relatively large distance this FSRQ is probably the most intensively studied blazar in every band of the electromagnetic spectrum.
Its SED has two broad bumps with the first
peak occurring at far-IR  ($\sim$ $10^{13}$ Hz) and the second one extending in
the MeV-GeV energy range. 3C 279 is highly variable at all frequencies of the
spectrum, particularly in the high frequency part of the two bumps, showing a
variability on time scales ranging from days to months.%
This behavior is usually interpreted assuming that the first peak is due to
the synchrotron emission of highly relativistic electrons accelerated in a jet
stemming from the nucleus, and the second peak generated by the Inverse
Compton (IC) emission of the same electrons interacting with low-energy
photons produced by the jet itself (Synchrotron Self Compton, SSC) or by
a region external to the jet (External Compton, EC). 

3C 279 is the first quasar that exhibited apparent
superluminal motion (Whitney et al. 1989) and high-resolution VLBI radio maps
show correlations between flare activity and components ejected from the core
(Wehrle et al. 2001). The observation of radio blobs emitted by 3C 279
(Lindfords et al. 2006) strongly supports the presence of a jet. It has been argued that the misalignment between the jet
and the line of sight is only two degrees (Lindfords et al. 2005).

3C 279 has been detected several times in the gamma-rays energy band by the EGRET
instrument on board $CGRO$ (Hartman et al. 2001) with integrated flux changes
up to a factor of 100.
\\The gamma-ray emission exhibited the largest amplitude variability on both long
(months) and short (days) time scales. On the basis of historical gamma-ray observations it is possible to
distinguish between high states (average flux equal or greater than about $10^{-6}$ ph
cm$^{-2}$ s$^{-1}$ above 100 MeV) and low states (average flux of the
order of $10^{-7}$ ph cm$^{-2}$ s$^{-1}$). 
The photon indices of the gamma-ray  
energy spectra during the different levels of activity of the source  
ranging from 1.7 to 2.4; a correlation between average fluxes and  
spectral indices is still debated 
(Nandikotkur et al. 2007; Hartman et al. 2001).

3C 279 was also detected by OSSE (50 keV--1
MeV) at the transition region from hard X-rays and gamma-rays and COMPTEL at
low energy (1--30 MeV) gamma-rays (McNaron-Brown et al. 1995). Although flux
variations were found in the OSSE and EGRET energy bands, only a marginal flux variations were
observed in the COMPTEL energy band. Hartman et al. (2001) argued that this
behaviour could be explained by considering the Comptonization of direct
accretion disc photons at COMPTEL energies and the Comptonization of accretion disc photons scattered into the jet region by the
Broad Line Region (BLR) clouds at EGRET energies.

Evidence for a thermal component due to the emission from the accretion disk
are noted by Pian et al. (1999). This Seyfert-like component,
present in other blazar as 3C 454.3 (Raiteri et al. 2007) and AO 0235+164
(Raiteri et al. 2006), is detected only in low activity states of the sources, when the contribution of the beamed synchrotron radiation is
less important. This could explain the smaller variability observed in the UV
than in optical band, under the assumption that the disk emission is not
variable on time scales less than a few years. 
Instead, in the optical band 3C 279 varies dramatically on different time scales
ranging from intense outbursts that last about one year to microvariability on
the scale of hours (see e.g. Kartaltepe and Balonek, 2007). Historically the
source has a R-band magnitude ranging between 12.5 and 16.5. The variability profile of
the flares observed seems to be consistent with the optical emission of 3C 279 being
dominated by synchrotron emission produced in the strong magnetic field of the
relativistic jet. 
\\

Recently the MAGIC
telescope has detected very high energy (VHE)
gamma-rays from 3C 279 (Albert et
al. 2008). The detection of the VHE gamma-ray emission from a source at such a
distance could constrain the current theories about the density of the
extragalactic background light (EBL), providing an indication that the
Universe appears more transparent at cosmological distance that believed.

In this Letter we present the analysis of the AGILE observation of 3C 279
between 9 July 2007 and 13 July 2007.

\begin{figure}[ht!]  
\label{mappa}
\begin{center}
\includegraphics[width=8cm]{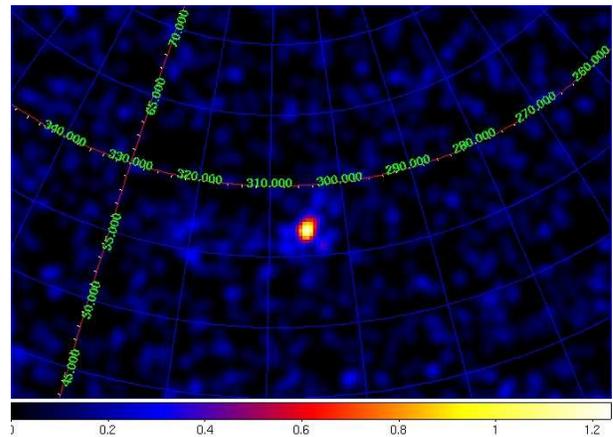}
\end{center}
\caption{Gaussian-smoothed counts map in Galactic coordinates for the 3C 279
  region over the observing period 9--13 July 2007. Only photons with energy greater than 100 MeV have been folded.}
\end{figure} 

\section{AGILE Data}

\subsection{AGILE Observation of 3C 279}

AGILE (Astrorivelatore Gamma a Immagini LEggero) is an Italian
Space Agency (ASI) mission (Tavani et al. 2008) devoted to high-energy
astrophysics studies.  

The satellite was successfully launched from the Indian base of Sriharikota on 23 April 2007 and has the peculiar characteristic to operate
simultaneously in the hard X-ray and gamma-ray energy bands yielding
broad-band coverage in the energy spectrum. The AGILE Instrument
combines the Gamma Ray Imaging Detector (GRID) sensitive in the energy range 30 MeV--50 GeV and the X-ray coded-mask
imager SuperAGILE (Feroci et al. 2007) sensitive in the energy range 18--60
keV. The GRID instrument is composed by a pair-production tracker based on the
silicon microstrip technology (Silicon Tracker; Prest et al. 2003, Barbiellini
et al. 2001), a calorimeter made of CsI bars (Mini-Calorimeter; Labanti et
al. 2006) placed below the Silicon Tracker and an Anti-coincidence System (ACS)
segmented in 13 panels of plastic scintillators (Perotti et al. 2006).

The Silicon Tracker and the on-board trigger logic are optimazed for gamma-ray
imaging in the 30 MeV--50 GeV energy band (Argan et al. 2004).

At the beginning of the Science Performance Verification Phase, AGILE repointed
the Virgo region and observed the blazar 3C 279 for a total of 44 hours of effective time. The
source was close (about 2 degrees) to the center of the field of view of the
Instrument for the
whole observing time. Super-AGILE observed 3C 279 for a total on-source net
exposure time of about 100 ks.

\subsection{Data Reduction and Analysis}

Level-1 AGILE-GRID data were analyzed using the AGILE Standard Analysis
Pipeline. The ACS and a set of hardware triggers perform the first reduction of the high rate of background events (charged particles and albedo gamma-rays) interacting with the instrument.
A dedicated software processes the signals coming from the GRID and the ACS,
performing a reconstruction and selection of events in order to  discriminate
between background events and gamma-rays, deriving for the latter the energy and direction of the incoming photons.
A simplified version of this software operate directly on board of AGILE in order to reduce 
the telemetry throughput
(Giuliani et al., 2006). 
A more complex version of the reconstruction and selection software is applied on ground, producing a photon list containing arrival time, energy and direction of every gamma-ray and the corresponding exposure maps.

Counts, exposure and Galactic background gamma-ray maps, the latter based on the diffuse emission model developed for
AGILE (Giuliani et al. 2004), were created with a bin-size of 0.$^{\circ}$25
$\times$ 0.$^{\circ}$25 for photons with energy greater than 100 MeV. We
selected only events flagged as confirmed gamma-ray events, and all events
collected during the South Atlantic Anomaly and whose reconstructed directions
form angles with the satellite-Earth vector smaller than 80$^{\circ}$ are rejected. 

In order to derive the average flux and spectrum of the source 
we run the AGILE Maximum Likelihood procedure (Chen et al. 2008) on the whole
observing period, according to Mattox et al. (1993). 
\begin{figure}[ht!]  
\includegraphics[width=9cm]{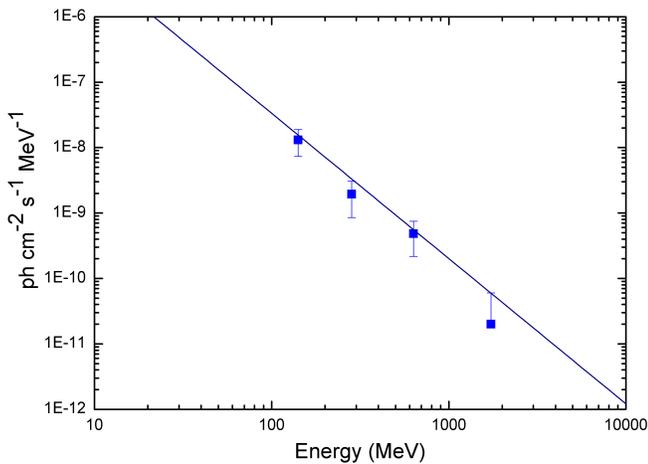}
\label{spec}
\caption{Gamma-ray photon spectrum for 3C 279 during the observation period 9
  -- 13 July 2007.%
Four energy bins were taken into account: 100--200 MeV, 200--400 MeV,
400--1000 MeV, 1000--3000 MeV. The solid line corresponds to a power law
function with photon index 2.22$\pm$0.23.}%
\end{figure}

\section{Optical and X-ray observations}

\subsection{Swift Observation}

The Swift X-ray Telescope (XRT; Burrows et al. 2004, 0.2--10 keV) data were
processed with standard procedures (xrtpipeline v0.11.6) adopting the standard
filtering and screening criteria. All the observations were carried in photon
counting (PC) mode and photons were selected with grades in the range 0--12.

Swift-XRT uncertainties are given at 90$\%$ confidence level for one
interesting parameter unless otherwise stated. Data were rebinned in
order to have at least 20 counts per energy bin and to use the $\chi^{2}$
statistics. 

Spectral analysis was performed using the Xspec fitting package 12.3.1,
with the results shown in Table 1. We fit the spectra with a 
absorbed power law with Galactic absorption  
fixed at the value $N_{H}$ = 2.05
$\times$ 10$^{20}$ cm$^{-2}$ (wabs*powerlaw model in Xspec). 
 
\begin{table}[h*]
\caption{Results of XRT observations of 3C 279. Power law model with $N_{\rm H}$ fixed to the
Galactic value of $2.05\times 10^{20}$ cm$^{-2}$. }
\begin{tabular}{cccc}
\hline
\multicolumn{1}{c}{Observation} &
\multicolumn{1}{c}{Flux 2-10 keV} &
 \multicolumn{1}{c}{Spectral slope} &
 \multicolumn{1}{c}{$ \chi^{2}_r$ (d.o.f.)} \\
 \multicolumn{1}{c}{date}&
 \multicolumn{1}{c}{erg cm$^{-2}$ s$^{-1}$} &
 \multicolumn{1}{c}{$\Gamma$}&
 \multicolumn{1}{c}{}\\
\hline
10-Jul-2007 & 1.20$\times 10^{-11}$ & $1.42 \pm 0.05$ &1.21 (73)\\
11-Jul-2007 & 1.17$\times 10^{-11}$ & $1.47 \pm 0.07$ &0.86 (52)\\
12-Jul-2007 & 1.05$\times 10^{-11}$ & $1.47 \pm 0.06$ &1.07 (57)\\
13-Jul-2007 & 1.13$\times 10^{-11}$ & $1.48 \pm 0.06$ &0.96 (50)\\
\hline
\end{tabular}
\label{tab.spectral_results}
\end{table}

\subsection{REM Observation}

The photometric optical observations were carried out
with the Rapid Eye Mount (REM, Zerbi et al. 2004), a robotic telescope located
at the ESO Cerro La Silla observatory (Chile). The REM telescope has a
Ritchey-Chretien configuration with a 60 cm f/2.2 primary and an overall f/8
focal ratio in a fast moving alt-azimuth mount providing two stable Nasmyth
focal stations. At one of the two foci the telescope simultaneously feeds, by
means of a dichroic, two cameras: REMIR for the NIR (Conconi et al. 2004) and
ROSS for the optical (Tosti et al. 2004), used in order to obtain nearly
simultaneous data. For a  
detailed description of the procedure of data reduction and analysis  
see e.g. Dolcini et al. 2007.

The telescope REM has continuously observed 3C 279 for about 1 year between
December 2006 and December 2007, including
the AGILE observation period. The light curve produced by REM in the R-band is shown in Figure 3.%

\begin{figure}[ht!]  
\includegraphics[width=8.5cm]{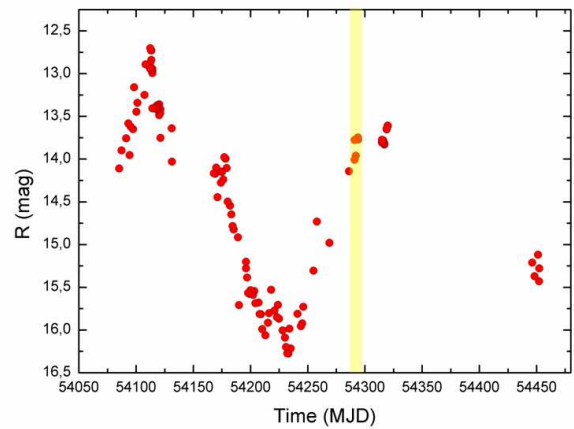}
\centering
\label{rem}
\caption{Long-term R-band light curve as observed by REM between December 2006
  and December 2007. The yellow shaded region indicates the period covered by the GRID observation.}
\end{figure}

\section{Results and Discussion}

During the period between 9 and 13 July 2007, AGILE detected
a gamma-ray source with position compatible with 3C 279 at a significance
level of 11.1 $\sigma$ (see Fig. 1) with an average flux of (210 $\pm$ 38) $\times$
10$^{-8}$ ph cm$^{-2}$ s$^{-1}$ for E $>$ 100 MeV, as derived from
a maximum likelihood analysis using radio position of the source (l =
305.10$^{\circ}$ b = 57.06$^{\circ}$).  
AGILE detected the source at a flux level comparable to that measured by EGRET
when the source was in a flaring state. 
Fitting the gamma-ray fluxes with a
constant model (the weighted mean of the 1-day average flux values) and following McLaughlin et al. (1996) we obtain a variability
coefficient of V = 0.32, indicating that the source
is not variable in gamma-ray band during the AGILE observation. 

Fig. 2 reports the average gamma-ray spectrum of 3C 279 for the AGILE observing
period. Fitting the data with a simple power law model we obtain a photon
index of $\Gamma$ = 2.22 $\pm$ 0.23. The photon index is calculated with the weighted
least squares method, considering only three energy bins for the fit: 100--200 MeV, 200--400 MeV and 400--1000 MeV. The source was not detected (above 5 $\sigma$) by the Super-AGILE Iterative
Removal of Sources (IROS) applied to the image, in the 20--60 keV energy
range. A 3 $\sigma$ upper limit of 10 mCrab was obtained from the observed count
rate by a study of the background fluctuations at the position of the source
and a simulation of the source and background contributions with
IROS. Instead, Swift-XRT detected the source with daily observation between
10 and 13 July at a flux nearly constant of about 10$^{-11}$ erg cm$^{-2}$
s$^{-1}$ (see Table 1). 

The multiwavelengh studies performed in the past on 3C 279 showed that during gamma ray flares most of the power emitted by the source lies in the high-energy gamma-ray band.
In accordance with leptonic models for blazars, the high-energy peak in the blazars SED is due to Inverse Compton emission from the relativistic electrons accelerated by the jet. 
In the case of 3C 279, the external Compton scattering of direct disk radiation (EDC) and external
Compton scattering of radiation from clouds (ECC) components dominate in high-energy gamma-rays yielding a total spectrum which varies as a function of the relative contribution of these two components. 
\\A possible correlation between gamma-rays flux value and spectral index is
subject of debate (Nandikotkur et al., 2007), but EGRET observations of 3C 279 hinted a gradual hardening
during the flaring states, that can be interpreted as the ECC component dominate during the flaring states. 
Only one flare (during EGRET observation P9) showed a soft spectrum EDC
dominated. Our AGILE observation seems to be similar to the P9 flare supporting the idea that 
a soft spectrum during flaring episodes is not a extremely rare event. 
Hartman et al. (2001) suggests that softening spectra can be due to a low state of the accretion disk occurred before of the EGRET observation P9, which led to a reduction of the ECC component.

From the R-band light curve (Fig. 3) emerges that a strong minimum occurred
about 2 months before the period covered by the GRID observations (indicated
by the yellow shaded region in figure).
This optical minimum might be correlated with a low accretion state of the
disk{; infact, even if the relation between highly relativistic jets and
  accretion processes in AGN is one of the fundamental open problems in
  astrophysics, the current theoretical models of the formation of jet suggest
that the power is generated by means of accretion, extracted from the disk
rotational energy and converted into kinetic power of the jet (Blandford $\&$
Payne 1982). 
If the accretion rate is related to the
jet power (see e.g. Ghisellini and Tavecchio 2008) then it is also related to the
synchrotron emission seen in the optical band. 
In this scenario, the reduction of activity of the disk causes} the decrease of the photon seed population produced by
the disk and then a ECC component deficit; this effect was delayed of two months (roughly the light travel time from the inner disk to
the Broad Line Region) coincident with the AGILE observation period. To test
this { hypothesis} we fit the optical, X-rays and gamma-rays
data with a SCC+EC model similar to the model used to fit the P9
EGRET observation in Hartman et al. { (2001) but with different parameters values}, finding a good agreement with the
data. We use a double power law distribution
for the electron's energy density with spectral index $p_1$ = 2.0 from
$\gamma_{min}$ = 100 to $\gamma_{break}$ = 600 and $p_2$ = 4.0
for $\gamma$ over 600, with a density at break $n_{e}$ = 30 cm$^{-3}$ { and
  $\gamma_{max}$ = 6 $\times$ 10$^{3}$. The blob have
radius $R$ = 2.5 $\times$ 10$^{16}$ cm, magnetic field B = 1.8 Gauss and it
moves with a bulk Lorentz factor $\Gamma$ = 13 at an angle $\theta$ = 2$^{\circ}$ with respect to the line of sight. The relativistic Doppler factor
is then $\delta$ = 21.5.} The accretion disk luminosity assumed is
$L_d$ = 5 $\times$ 10$^{45}$ erg s$^{-1}$ with a Broad Line Region
reprocessing a 10$\%$ of the illuminating continuum.
Fig. 4 shows the SED for the GRID observing
period, including simultaneous optical (REM) and X-ray (Swift) data { and
indicating individual model components.}

\begin{figure}[ht!]  
\includegraphics[width=10.0cm]{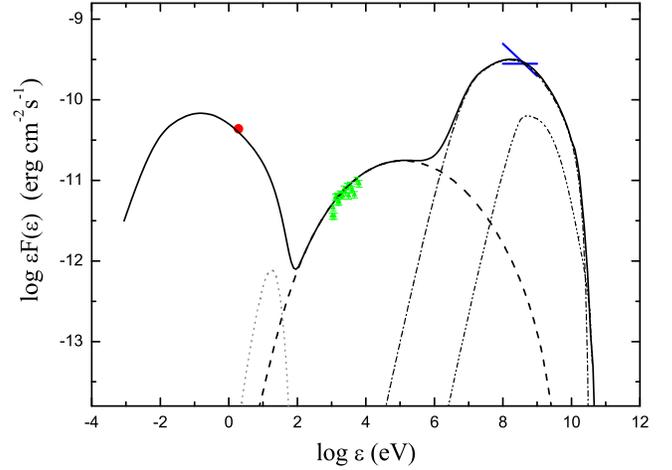}
\centering
\label{sed}
\caption{Spectral Energy Distribution of 3C 279 for the AGILE-GRID
  observation, including simultaneous optical (REM) and X-ray (Swift)
  data. {{ The dotted, dashed, dot-dashed and double-dot dashed lines
  represent the contributions of the accretion disk blackbody, the SSC, the external
  Compton on the disk radiation and the external Compton on the Broad Line Region
  radiation, respectevely.}}}
\end{figure}

\section{Conclusions}

During the period 2007 June 9-13 AGILE detected a highly significant
source (11.1 sigma for E$>$ 100 MeV) associated to the FSRQ 3C 279.
The source had an average  flux of (210 $\pm$ 38) $\times$ 10$^{-8}$  ph cm$^{-2}$ s$^{-1}$ over the period covered by AGILE observation.
Gamma-ray flux variations on 1-day time scale seems not to be present in our data, while
longer timescales cannot be investigated due to the short observing time.
The spectra measured by AGILE is compatible with a power law distribution of
slope 2.22$\pm$0.23 in the energy range 100 MeV - 1 GeV. This soft spectrum observed
during a flaring episode, already observed during EGRET P9 observation, could
be an indication of a dominant contribution of the EDC emission compared to ECC emission. 
The Inverse Compton scattering of relativistic electrons of synchrotron or ambient
photons is responsible for the emission at hard X-rays and gamma-ray
energies for FSRQ as 3C 279. Claryfing the exact nature of the seed photons
for the IC scattering
would explain the origin of the huge amplitude variations exhibited by 3C 279
at the highest energies. 

\begin{acknowledgements}
The AGILE Mission is funded by the Italian Space Agency (ASI) with
scientific and programmatic participation by the Italian Institute
of Astrophysics (INAF) and the Italian Institute of Nuclear
Physics (INFN). We wish to express our gratitude to
the Carlo Gavazzi Space, Thales Alenia Space, Telespazio and ASDC/Dataspazio
Teams that implemented the necessary procedures to carry out the
AGILE re-pointing. {\it Facilities:AGILE}
\end{acknowledgements}

\end{document}